\documentclass[11pt, a4paper]{article}

\usepackage{amsmath}
\usepackage{amsfonts}
\usepackage{amssymb}
\usepackage{graphicx, rotating}
\usepackage{epstopdf}
\usepackage{epsfig}
\usepackage{latexsym}
\usepackage{graphicx}
\usepackage{color}
\usepackage{amsmath,bm,amssymb}
\usepackage{cite}
\usepackage{slashed}
\usepackage{hyperref}
\usepackage{comment}
\usepackage[utf8]{inputenc}
\usepackage{soul}
\hypersetup{colorlinks, citecolor=bluscuro, linkcolor=black, urlcolor=bluscuro}
\definecolor{rossos}{cmyk}{0,1,1,0.55}
\definecolor{bluscuro}{rgb}{0.15, 0.2, .85}
\definecolor{bluchiaro}{cmyk}{1,.3,0.,0.1}

\newcommand{\be}{\begin{equation}}
\newcommand{\ee}{\end{equation}}
\newcommand{\bea}{\begin{eqnarray}}
\newcommand{\eea}{\end{eqnarray}}

\newcommand{\GeV}{\,\mathrm{GeV}}

 \def\bea{\begin{eqnarray}}
  \def\eea{\end{eqnarray}}

\newcommand{\mH}{m_{H^0}}
\newcommand{\mA}{m_{A}}
\newcommand{\mHc}{m_{H^+}}
\newcommand{\tanb}{\tan\beta}

\newcommand{\like}{{\cal L}}

\newcommand{\relic}{\Omega_\mathrm{DM} h^2}
\newcommand{\Omegaobs}{\Omega_{\rm DM}^\mathrm{obs}}
\newcommand{\Omegachi}{\Omega_{\chi}}

\begin{document}




\begin{titlepage}
\begin{flushright}
IFT-UAM/CSIC-20-161
\end{flushright}
\begin{center} ~~\\
\vspace{0.5cm} 
\Large { \bf\Large 2HDM singlet portal to dark matter} 
\vspace*{1.5cm}

\normalsize{
{\bf  M. E. Cabrera$^a$\footnote[1]{meugenia@ecfm.usac.edu.gt},
J. A. Casas$^{b}$\footnote[2]{alberto.casas@uam.es},
 A. Delgado$^c$\footnote[3]{antonio.delgado@nd.edu}
 and S. Robles$^d$\footnote[4]{sandra.robles@unimelb.edu.au}} \\

\smallskip  \medskip
$^a$\emph{Instituto de Investigaci\'on en Ciencias F\'isicas y Matem\'aticas, \\ Universidad de San Carlos de Guatemala, Ciudad Universitaria, 01012 Guatemala, Guatemala}\\
$^b$\emph{Instituto de F\'\i sica Te\'orica, IFT-UAM/CSIC,}\\
\emph{Universidad Aut\'onoma de Madrid, Cantoblanco, 28049 Madrid, Spain}\\
$^c${\it Department of Physics, University of Notre Dame,}\\
{\it Notre Dame, IN 46556, USA}}\\
$^d$\emph{ARC Centre of Excellence for Dark Matter Particle Physics, \\
School of Physics, The University of Melbourne, Victoria 3010, Australia}

\medskip

\vskip0.6in 

\end{center}

\centerline{ \large\bf Abstract}
\vspace{.5cm}
\noindent

Higgs portal models are the most minimal way to explain the relic abundance of the Universe. They add just a singlet that only couples to the Higgs through a single parameter that controls both the dark matter relic abundance and the direct detection cross-section. Unfortunately this scenario, either with scalar or fermionic dark matter, is almost ruled out by the latter. In this paper we analyze the Higgs-portal idea with fermionic dark matter in the context of a 2HDM. By disentangling the couplings responsible for the correct relic density from those that control the direct detection cross section we are able to open the parameter space and find wide regions consistent with both the observed relic density and all the current bounds.

\vspace*{2mm}
\end{titlepage}


\section{Introduction}
\label{sec:Intro}

Dark matter (DM) is the most  established reason for physics beyond the Standard Model (SM). Higgs-portal models of DM~\cite{Burgess:2000yq,Kim:2006af}, where the annihilation of the DM particles occurs thanks to their interactions with the Higgs, are among the most popular and economical DM scenarios. Unfortunately, the simplest version of this framework, consisting of the SM content plus a neutral particle coupled to the Higgs sector, is essentially excluded in the whole parameter space, except for very large DM masses and inside a narrow
region around the Higgs-funnel $(m_{\rm DM}\sim m_h/2)$. The main responsible for this exclusion are the constraints from DM direct detection (DD)~\cite{Aprile:2018dbl}. They are particularly stringent because, in such scenario, the required  DM annihilation  in the early Universe and the DD cross sections are controlled by the same coupling.

This situation invites to consider (economical) extensions of the Higgs portal which could circumvent the current DD constraints. 
In this paper, we consider the scenario, where the DM sector still consists of a single particle (a neutral fermion), but the Higgs sector comprises two Higgs doublets, i.e. the well-known 2HDM. 
This allows to break the previous connection between the relic abundance and the DD cross section, opening vast regions of the parameter consistent with both constraints.
Previous works in similar directions have been carried out in refs.~\cite{Berlin:2015wwa,Bell:2016ekl,Bell:2017rgi,Arcadi:2017wqi,Baum:2017enm,Bauer:2017fsw}.

The outline of this paper is as follows: In section \ref{sec:model} we review the 2HDM and write the most general Lagrangian coupled to a fermion singlet, which has the form of an Effective Field Theory (EFT) with four independent couplings. In section \ref{sec:DMpheno} we describe the main aspects of the DM phenomenology in such scenario, focusing on DM-annihilation and DD processes. In section \ref{sec:DMScan} we scan the parameter space using representative benchmarks and discuss the results. As mentioned above, very large regions of the parameter space are now rescued. We compare the results with those of the conventional Higgs-portal, which we up-date. The implications and prospects for indirect detection are also analyzed and discussed. Finally, in section \ref{sec:Conclusions} we summarize our conclusions.

\section{The 2HDM-portal scenario}
\label{sec:model}

We consider a 2HDM scenario together with a dark sector which contains just a neutral fermion $\chi$, which for minimality we consider to be Majorana. The interactions of $\chi$ with the Standard Model (SM) fields are necessarily non-renormalizable  which can be interpreted as an effective theory. The most general lowest-dimensional Lagrangian is then given by
\be
-{\cal L}_\chi \supset \frac{1}{2}M\bar \chi\chi +\frac{1}{2\Lambda}\bar \chi\chi\left(
\hat y_1|\Phi_1|^2 +\hat y_2|\Phi_2|^2 +(\hat y_3\bar \Phi_1 \Phi_2 + {\rm h.c.})
\right)
+ 
\frac{1}{2\Lambda}\bar \chi\gamma_5\chi\left(
\hat y_4\bar \Phi_1 \Phi_2 + {\rm h.c.}
\right),
\label{L1}
\ee
where $\Phi_1$ and $\Phi_2$ are the Higgses, both with hypercharge $1/2$, and 
$\bar \Phi_{1,2}= i\sigma_2 \Phi_{1,2}^*$. $\Lambda$ denotes the scale of new physics giving rise to the above effective operators. A possible UV completion of the above Lagrangian is to include a fermion doublet of mass $\Lambda$ with appropriate couplings.

A correct electroweak symmetry breaking requires the neutral part of the Higgs doublets to acquire vacuum expectation values (VEVs), $v_1, v_2$, such that $v_1^2+v_2^2=v^2= (246 \ {\rm GeV})^2$. This in turn implies an appropriate structure of the 2HDM Higgs potential, which we do not show explicitly here (for further details see ref.~\cite{Bernon:2015qea}).
As usual, we define the $\beta$ angle so that
\be
v_1=v\ \cos\beta, \qquad v_2= v\  \sin\beta .
\ee
The actual value of $\beta$ depends on the parameters of the 2HDM potential.

From now on we will consider the Higgses in the mass (eigenstate) basis, $\{H_1, H_2\}$, related to the initial one by
\be
\left(\begin{array}{c} \Phi_2\\ \Phi_1 \end{array}   \right) = 
 \left(\begin{array}{cc} \cos\alpha & \sin\alpha \\ 
-\sin\alpha & \cos\alpha  \end{array}\right)
\left(\begin{array}{c} H_1\\H_2  \end{array}   \right),
\label{alpha_rot} 
\ee
where $\alpha$ is determined by the parameters of the 2HDM potential.
In the mass basis the Lagrangian Eq.~(\ref{L1}) takes a completely analogous form
\be
-{\cal L}_{\rm fer} \supset \frac{1}{2}M\bar \chi\chi +\frac{1}{2\Lambda}\bar \chi\chi\left(
y_1|H_1|^2 +y_2|H_2|^2 +(y_3\bar H_1 H_2 + {\rm h.c.})
\right) + 
\frac{1}{2\Lambda}\bar \chi\gamma_5\chi\left(
y_4\bar H_1 H_2 - {\rm h.c.}\right),
\label{L3}
\ee
where the relation between the $y_{i}$ and $\hat y_{i}$ couplings is straightforward from Eq.~(\ref{alpha_rot}).
In principle all the parameters in the previous Lagrangian are complex, but from field re-definitions there are only three independent phases, namely  $M y_1^*$, $M y_2^*$ and $y_3 y_4^*$, so we can assume $M\geq 0$.

\vspace{0.3cm}

As it is well known, experimental observations require the 125 GeV Higgs boson to have almost the same properties as that of the SM \cite{Aad:2012tfa,Chatrchyan:2012ufa,Aad:2013wqa,Chatrchyan:2013lba,Sirunyan:2017exp,Aaboud:2018wps}. Hence a conservative approach, which we will adopt throughout the paper, is that such identification is exact, which is known as the ``alignment limit". Then, assuming that the light Higgs, $h^0$, plays the role of the SM Higgs boson, the alignment limit occurs when $c_{\beta-\alpha}=0$, i.e. $\alpha=\beta-\pi/2$. Equivalently, this means that the heavy Higgs does not obtain a VEV. We will discuss in section~\ref{sec:DMScan} the two main instances where the alignment takes place.

Now, in this limit we can write
\be
H_1=
\left(\begin{array}{c} G^+\\\frac{1}{\sqrt{2}}(v+h^0 +iG^0) \end{array}   \right),\qquad
H_2=
\left(\begin{array}{c} H^+\\\frac{1}{\sqrt{2}}(H^0 +iA) \end{array}   \right),
\label{mbasis}
\ee
where $G$ are Goldstone bosons and $h^0$ ($H^0$) are the SM (heavy) neutral Higgs states. 
It is straightforward to check that, when plugging Eq.~(\ref{L3}) in the Lagrangian, the couplings 
 $y_3$ and $y_4$, if complex, lead to CP-violating interactions, which are subject to restrictions \cite{Tuzon:2010vt,Jung:2010ik,Enomoto:2015wbn}; so we will assume they are real. Then, upon electroweak symmetry breaking, they induce trilinear couplings in the form
\be
-{\cal L}_{\rm tril} \supset \frac{v}{2\Lambda}\bar \chi\chi\left(
y_1 h^0 + y_3 H^0 \right)
+ y_4
\frac{v}{2\Lambda}
i \bar \chi\gamma_5\chi\ A
+ {\rm h.c.}
\ee
These interactions will control the different funnels. On top of that there are quartic interactions that are going to be important for the different channels as we will explain in the next section.

\section{DM phenomenology}
\label{sec:DMpheno}

The interactions in the Lagrangian, Eq.~(\ref{L3}), lead to a number of DM annihilation channels. Their efficiencies depend on the relevant $y_i$ coupling in each case. Namely,
\bea
&y_1&:\hspace{0.3cm} \chi\chi\rightarrow h^0 \rightarrow {\rm SM\ SM}, \ \
\chi\chi\rightarrow h^0h^0,\ \ \chi\chi\rightarrow ZZ , \ \ \chi\chi\rightarrow W^+W^- ,
\nonumber\\
&y_2&:\hspace{0.3cm} \chi\chi\rightarrow H^0H^0,\ \ \chi\chi\rightarrow AA , \ \ \chi\chi\rightarrow H^+H^- ,
\nonumber\\
&y_3&:\hspace{0.3cm} \chi\chi\rightarrow H^0 \rightarrow {\rm SM\ SM}, \ \ \chi\chi\rightarrow h^0H^0,\ \ \chi\chi\rightarrow ZA , \ \ \chi\chi\rightarrow W^\pm H^\mp ,
\nonumber\\
&y_4&:\hspace{0.3cm} \chi\chi\rightarrow A \rightarrow {\rm SM\ SM}, \ \ \chi\chi\rightarrow h^0A,\ \ \chi\chi\rightarrow H^0A,\ \  \chi\chi\rightarrow ZH^0 , \ \ \chi\chi\rightarrow W^\pm H^\mp .
\label{annihilations}
\eea
We immediately see the presence of resonant annihilation for $m_h/2, m_{H^0}/2, m_A/2$, i.e. the well known $h^0, H^0, A$ funnels; but clearly there are many other possibilities of annihilation.

Concerning DD processes, only the $y_1$ and $y_3$ couplings lead to dangerous scattering cross sections, as they induce tree-level DD amplitudes, $\chi q \rightarrow \chi q$, with a neutral Higgs ($h^0$ and $H^0$ respectively) propagating in the $t-$channel. In contrast, $y_4$ induces DD processes mediated by the pseudoscalar $A$, and thus momentum suppressed (and spin-dependent), while $y_2$ does not lead to any DD tree-level process. Thus, compared to the SM Higgs sector, in the 2HDM portal scenario it is much easier to have suitable DM annihilation without paying a DD price. Hence one expects large allowed regions in the parameter space. Actually, even for the annihilation processes triggered by $y_1$, $y_3$, it may happen that the  corresponding contributions to DD cancel out, leading to the so-called blind-spot solutions.

Regarding the last point, the amplitude for the spin-independent  DM-nucleon scattering, $\chi N \rightarrow \chi N$, mediated by a Higgs ($h^0$ or $H^0$ in the $t-$channel) is essentially proportional to
\be
B_N\equiv \sum_q \left[y_{1} +  \frac{m_h^2}{m_H^2} C_q\  y_{3}\right] f_q^N,
\label{BN}
\ee
where $q$ runs over the quarks inside the nucleon and   $f_q^N$ (with $N=p,n$) are the hadronic matrix elements (related to the mass fraction of $q$ inside the proton or the neutron).  Finally, the factor $C_q$ denotes the ratio of the heavy Higgs-quark coupling, $H^0qq$, to that of the SM Higgs, $h^0qq$.

The $C_q$ factors depend both on the type of the quark $q$ considered and on the type of 2HDM at hand. They are given in Table \ref{tab:couplings} for the Type I and Type II 2HDMs, together with the analogous factors for the $Aqq$ couplings. There exist two additional 2HDMs, which are also safe regarding flavor changing neutral currents (FCNC): the so-called X (or “lepton-specific”) and Y (or “flipped”) models. They have the same $C_q$ factors as the Type I and Type II, respectively.
\begin{table}[t]
    \centering
    \begin{tabular}{|c||c|c||c|c|}
    \hline
    \multicolumn{1}{|c||}{} &
    \multicolumn{2}{|c||}{Type I} & \multicolumn{2}{|c|}{Type II}\\
    \hline
      Higgs &
      $u-$quarks & $d-$quarks and leptons&  $u-$quarks & $d-$quarks and leptons \\
      \hline
      $h^0$ &	$\cos\alpha/\sin\beta$ & $\cos\alpha/\sin\beta$ & $\cos\alpha/\sin\beta$ & $-\sin\alpha/\cos\beta$\\
      $H^0$ &	$\sin\alpha/\sin\beta$ & $\sin\alpha/\sin\beta$ & $\sin\alpha/\sin\beta$ & $\cos\alpha/\cos\beta$\\
      $A$ &	$\cot\beta$ & $-\cot\beta$ & $\cot\beta$ & $\tan\beta$\\
      \hline     
    \end{tabular}
    \caption{Ratios of the couplings of the Higgs states to fermions in  the Type I and Type II 2HDMs with respect to those of the SM. }
    \label{tab:couplings}
\end{table}

Note that in the alignment limit, $\alpha=\beta-\pi/2$, so we recover the SM couplings of the ordinary Higgs, $h^0$. In contrast, the couplings of the heavy Higgs $H^0$ to quarks acquire factors
\bea
&C_u=-\cot \beta,&\qquad C_d=-\cot\beta \qquad  \ \text{(Type I)}, 
\nonumber\\
&C_u=-\cot \beta,&\qquad C_d=\tan\beta \qquad \quad \text{(Type II)},
\eea
to be plugged in Eq.~(\ref{BN}) for the blind spot. It is worth noticing that the blind spot condition, $B_N=0$, cannot be simultaneously fulfilled for protons and neutrons in an exact way.

\section{Scan of the model and results}
\label{sec:DMScan}

\begin{figure}[t]
    \centering
    \includegraphics[scale=0.6]{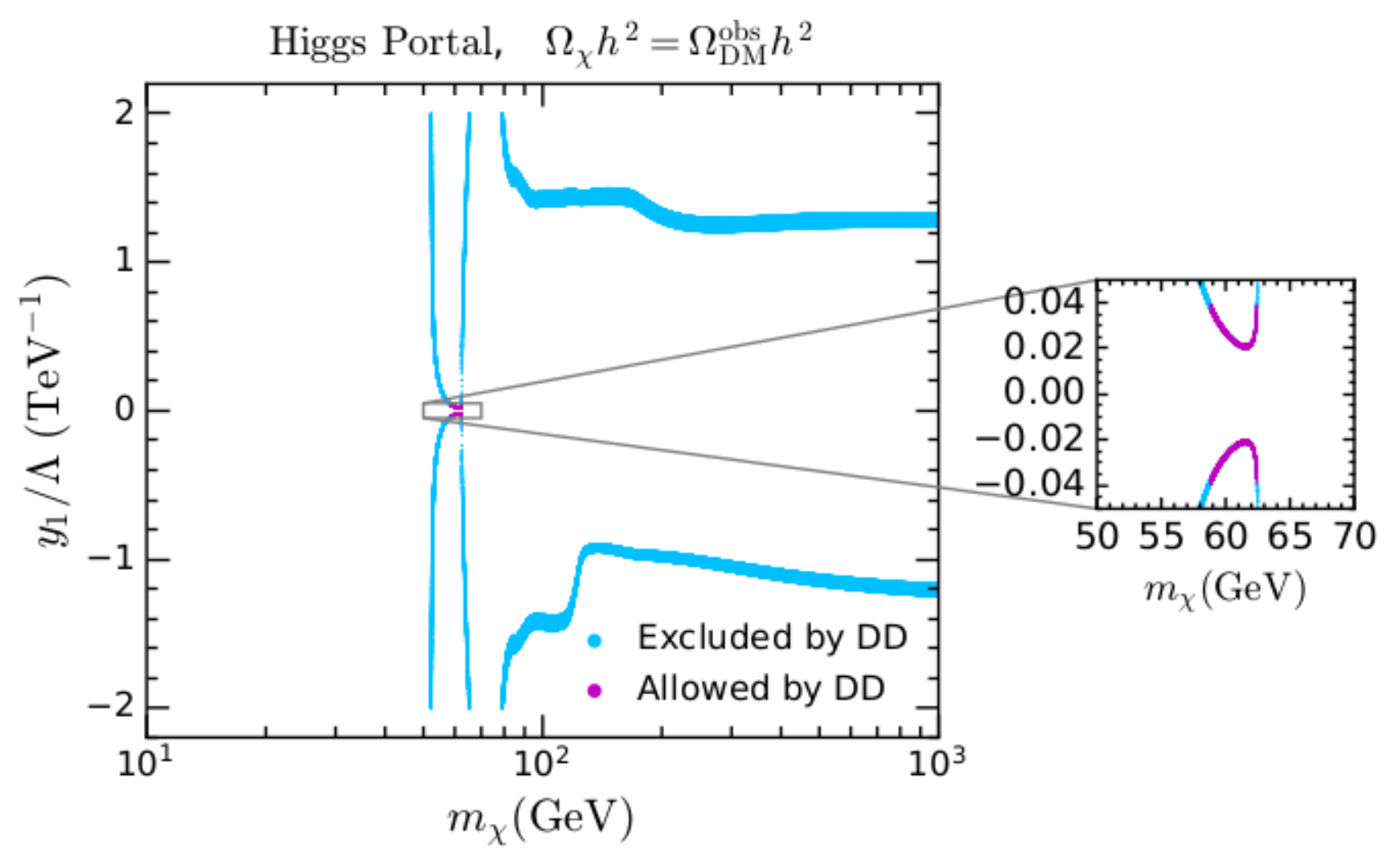}
    \caption{Regions of the parameter space in the decoupling limit that fulfill the constraint on the DM relic abundance $\Omegachi h^2=\Omegaobs h^2$. Points that also satisfy the XENON1T bound are shown in magenta and in the inset plot.}
    \label{fig:decoupling}
\end{figure}

Before performing a complete scan of the model, we consider the trivial case in which the alignment is achieved via decoupling, i.e. $m_{H^0}, m_{H^\pm}, m_{A} \gg m_{h^0}$. Then at low energy the Higgs sector is indistinguishable from that of the SM, so we recover the old Higgs-portal scenario with $y_1$ as the only relevant coupling. It is however interesting to show the results in this limit for later comparison with the case in which the alignment occurs without decoupling. The allowed region of the parameter space is shown in Fig.~\ref{fig:decoupling}. As mentioned in section \ref{sec:Intro}, only a narrow band at the Higgs funnel, plus the region of very large DM masses, survive.

As already mentioned, the main reason for this behavior is that a single coupling, $y_1$, controls all the relevant processes; and outside the Higgs-funnel region the required size of $y_1$ to have appropriate DM annihilation leads to a too-large (spin-independent) DD cross section. Unfortunately, this virtually excludes the conventional Higgs-portal as a viable scenario.

Let us now go to the more interesting case of alignment without decoupling.
This happens when the $Z_6$ coupling in the Higgs-potential (see ref.~\cite{Bernon:2015qea}) is vanishing or very small. Then, the alignment condition, $\alpha=\beta-\pi/2$, is satisfied for any value of $\mH, m_{H^\pm}, m_A$. Of course, the latter are subject to experimental constraints, which depend on the 2HDM considered. For the Type I  the bounds are quite mild, in particular $\mH$ can be close to $m_h=125$ GeV without conflict with observations. 

On the other hand, for the Type II the bounds are more restrictive, essentially because $H^0\to \tau\tau$ is enhanced by $\tan\beta$ (see Table \ref{tab:couplings}). A safe bound for the heavy Higgs is $\mH\geq 400$ GeV, although it can be much lighter if $m_A\geq 400$ GeV \cite{Bernon:2015qea}.

 \begin{table}[tb]
    \centering
    \begin{tabular}{|c|c|c|c||c|c|c|c|}
    \hline
    \multicolumn{4}{|c||}{Type I} & \multicolumn{4}{|c|}{Type II}\\
    \hline
      $\mH(\GeV)$ & $\mA(\GeV)$& $\mHc(\GeV)$  & $\tanb$ &  $\mH(\GeV)$ & $\mA(\GeV)$ & $\mHc(\GeV)$  & $\tanb$\\
      \hline
      300 & 300 & 300 & 5 & 300 & 300 & 300 & 5 \\
      500 & 300 & 500 & 5 & 300 & 500 & 500 & 30\\
      700 & 700 & 700 & 5 & 700 & 700 & 700 & 5\\
      \hline     
    \end{tabular}
    \caption{Benchmarks for the Type I and Type II 2HDMs.}
    \label{tab:benchmarks}
\end{table}

These aspects lead us to consider the benchmark models given in Table~\ref{tab:benchmarks}. In order to analyze them, we have  implemented the model in {\tt FeynRules}~\cite{Christensen:2008py,Alloul:2013bka}, interfaced with {\tt CalcHEP}~\cite{Christensen:2009jx}, taking the publicly available 2HDM model files~\cite{Degrande:2014vpa} as a starting point. The parameters of the model are those appearing  in the fermionic Lagrangian, Eq.~(\ref{L3}), plus those involved in the scalar potential. The latter, in particular, determine the masses of the Higgs particles and the $\alpha$ and $\beta$ angles, which are the relevant quantities for our study. On the other hand, the Higgs quartic couplings play no role, so we have used the default values in the FeynRules implementation.
Hence the parameters of our analysis are
\be
\{m_{h^0},m_{A},m_{H^0},m_{H^\pm},\alpha,\tan\beta;\frac{y_i}{\Lambda},M\}.
\label{params}
\ee
where $i=1,..., 4$, and, in the alignment limit, $\alpha = \beta-\pi/2$.

For every benchmark, the masses of the heavy Higgses and the value of $\tan\beta$ are taken from Table~\ref{tab:benchmarks}. Then, without loss of generality, we set $\Lambda=2$ TeV, while the remaining parameters are varied in the ranges
\be
|y_i|<4, \ i=1,\cdots,4, \qquad M \in [10,1000] \GeV,
\label{eq:params}
\ee
Since we can choose $y_4\geq 0$ (see discussion below Eq.~\ref{L3}), this means that for every benchmark we explore $y_i/\Lambda$ in the range $[-2,2]$ TeV$^{-1}$ for $i=1,2,3$ and $[0,2]$ TeV$^{-1}$ for  $i=4$.

At each point of the examined parameter-space, the DM  relic abundance and DM-nucleon elastic scattering cross section have been calculated with the {\tt microOMEGAS 5.0.8} package \cite{Belanger:2006is}. 
The exploration of the parameter space have been performed with {\tt MultiNest}~\cite{Feroz:2007kg,Feroz:2008xx,Feroz:2013hea}, a code based on the computation of the posterior probability density function. Even though we are not interested in the posterior, the code is a very efficient tool to scan the parameter space.
The accuracy of the scan depends on the required precision of the evidence, meaning that a judicious choice of the priors is important for a complete
exploration.
We have considered logarithmic priors on all the parameters in Eq.~(\ref{eq:params}). Besides, the joint likelihood is constructed  as
\begin{equation}
 \log \like_{\rm Joint} ~=~ \log\like_{\relic} + \log\like_{\rm Xenon1T}~, 
 \label{eq:jointlike}
\end{equation}
where $\like_{\relic}$ is implemented as a Gaussian distribution centered at the observed value \cite{Aghanim:2018eyx} and $\like_{\rm Xenon1T}$ is calculated using {\tt RAPIDD} \cite{Cerdeno:2018bty},  tuned to the latest XENON1T results \cite{Aprile:2018dbl,Aprile:2019dbj}. 

\begin{figure}[t]
    \centering
    \includegraphics[width=0.775\textwidth]{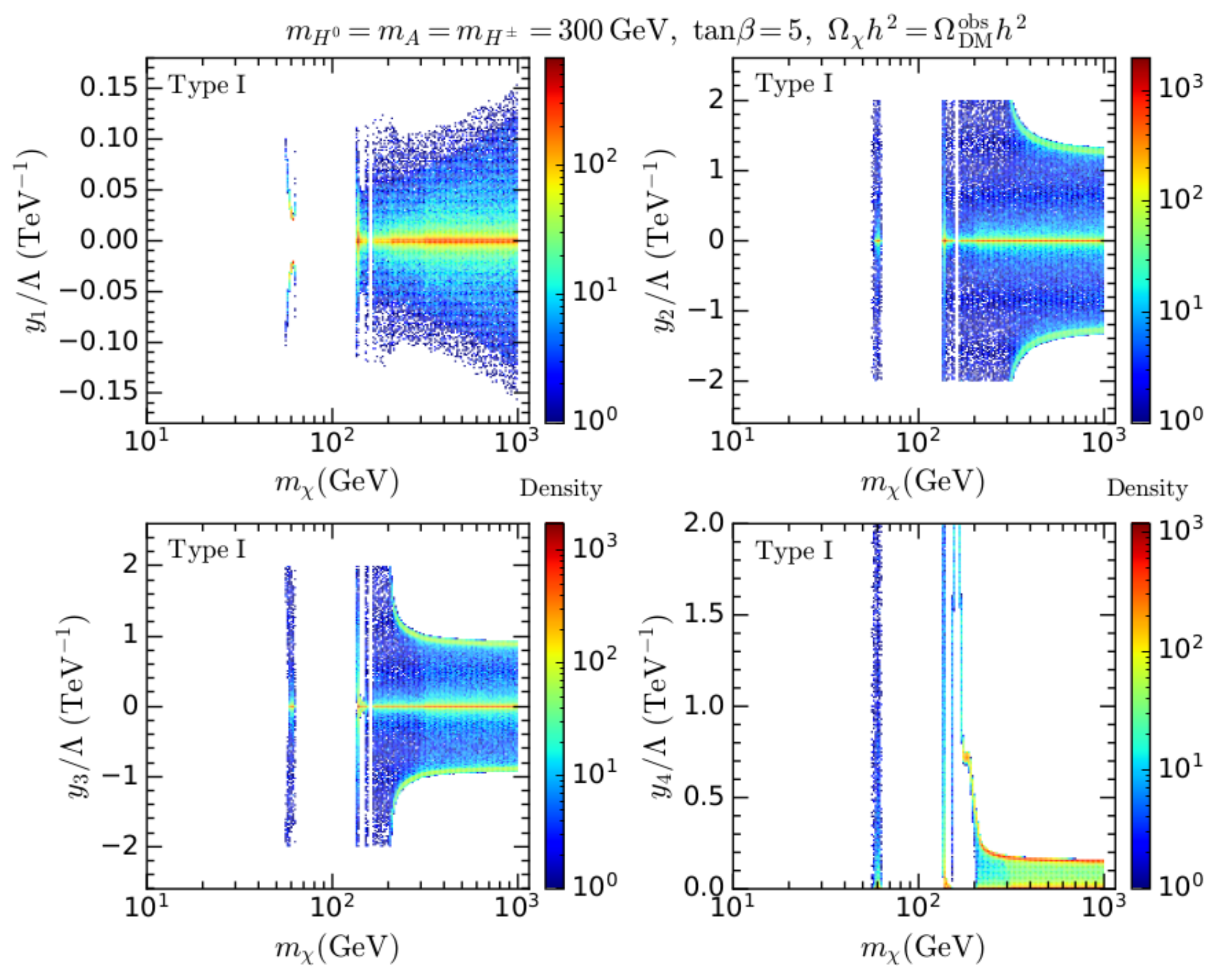}
    \caption{Regions of the parameter space that fulfil both the constraint on the DM relic abundance and direct detection limits for the Type I 2HDM with $\tan\beta=5$ and $\mH=\mA=\mHc=300\GeV$ (benchmark 1 in Table~\ref{tab:benchmarks}). The color bar denotes the relative density of points.}
    \label{fig:type1_bchm1}
\end{figure}

Fig.~\ref{fig:type1_bchm1}  shows the allowed values of the four $y_i$ couplings and the dark matter mass, projected on the
$m_\chi-y_i$ planes, for the first benchmark of the Type I 2HDM, see Table~\ref{tab:benchmarks}. The plots, in particular that  corresponding to the $y_1$ coupling, are to be compared with the conventional Higgs-portal, shown in Fig.~\ref{fig:decoupling}. As expected, the range of allowed values for the DM mass and the couplings are now much wider. This comes from the fact that, for a 2HDM, DM annihilation and DD processes are not controlled by the same couplings or combinations of couplings, as discussed in section~\ref{sec:DMpheno}. In the figure it is visible the Higgs funnel at masses $m_\chi\sim 60$ GeV, where $y_{2,3,4}$ can essentially take any value since they do not control the coupling of DM to the Higgs. Actually, there is a small correlation between the values of $y_1$ and $y_3$, corresponding to the points rescued by an approximate blind spot condition, see Eq.~(\ref{BN}). 
Then, there is  a gap until $m_\chi\sim150$ GeV, where the  heavy Higgs and pseudoscalar funnels appear, controlled by $y_3$ and $y_4$ respectively. Their V-shape is visible, especially in the plot for $y_4$.
Next, there is another small gap up to $m_\chi\sim 170$ GeV, where the $t\overline{t}$ channel for DM annihilation opens. Notice that in this region the allowed values of $y_4$ are more precise than those of the other couplings. This is because the main annihilation process occurs through a pseudoscalar in the $s-$channel and is thus controlled by $y_4$. The reason for this dominance is that, contrary to $y_3$, large values of $y_4$ are not dangerous for DD, since the corresponding cross section is spin-dependent and momentum-suppressed.
Moreover, the contribution of the interactions mediated by the pseudoscalar to the annihilation cross section is larger than those mediated by the heavy Higgs, due to kinematical factors. Therefore $y_4$ is more constrained from above by the relic density. Beyond that point  many new channels are opened, see Eq.~(\ref{annihilations}), so the allowed region of the parameter space enters a continuum. Notice that for $m_\chi>300$ GeV, the density of allowed points increases substantially. This is due to the annihilation processes in two heavy Higgs states driven by $y_2$, see Eq.~(\ref{annihilations}). These processes occur without paying any “DD price”, as $y_2$ does not drive any tree-level DM-nucleon scattering. 
It is also worth noting that the allowed values for $y_1$ are substantially smaller than for the other couplings, since $y_1$ is the most dangerous coupling regarding DD bounds. In contrast, the values of $y_2$ are the least restricted. On the other hand, the allowed values of $y_3$ are larger that those of $y_4$, due to the above mentioned enhancement in pseudoscalar mediated annihilation channels, compared to those mediated by the heavy CP-even Higgs.

\begin{figure}[h]
    \centering
    \includegraphics[width=0.725\textwidth]{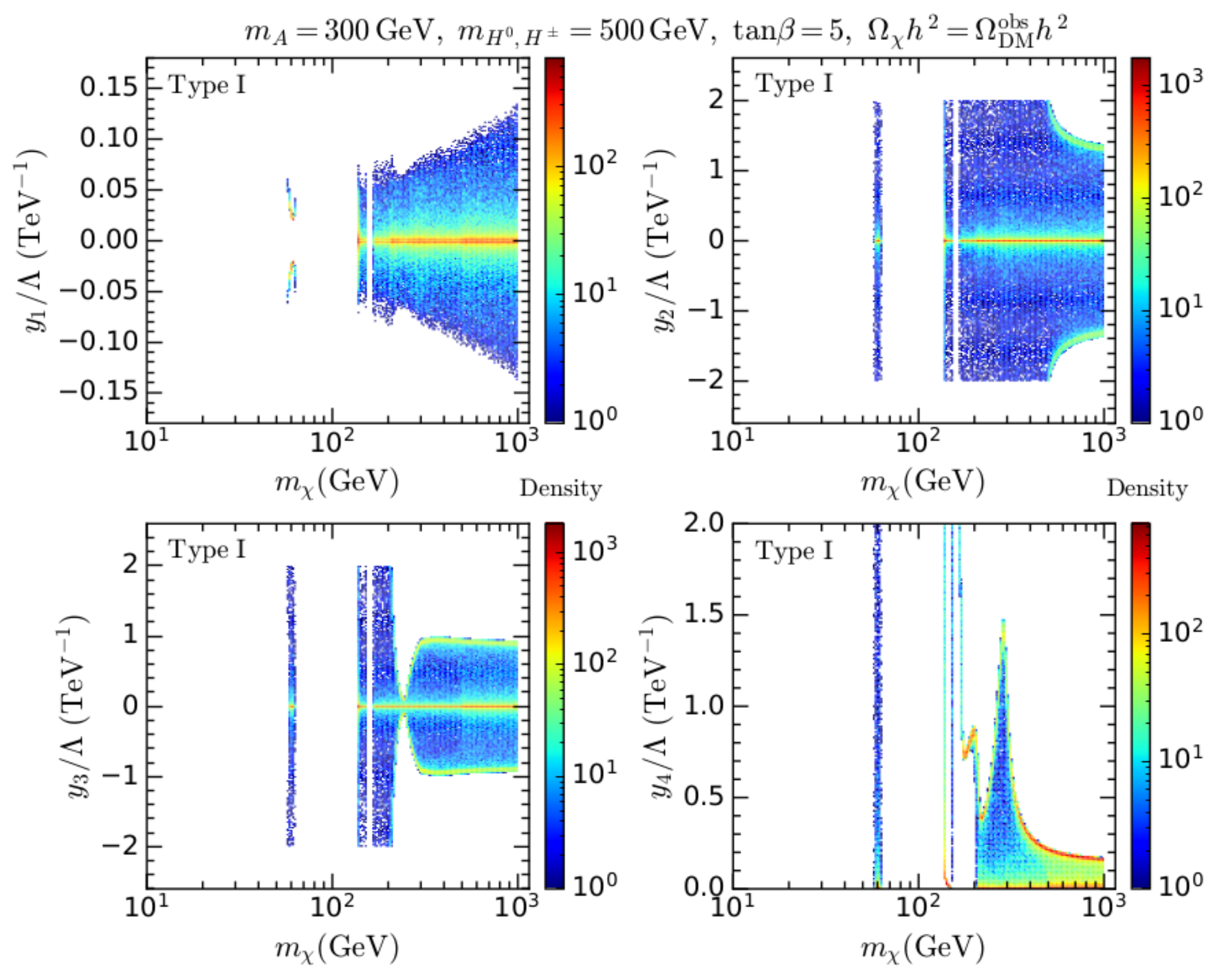}
    \caption{The same as Fig.~\ref{fig:type1_bchm1} for the Type I 2HDM with $\tan\beta=5$, $\mH=\mHc=500\GeV$ and $\mA=300\GeV$  (benchmark 2 in Table~\ref{tab:benchmarks}).}
    \label{fig:type1_bchm2}
\end{figure}

\begin{figure}[h]
    \centering
    \includegraphics[width=0.725\textwidth]{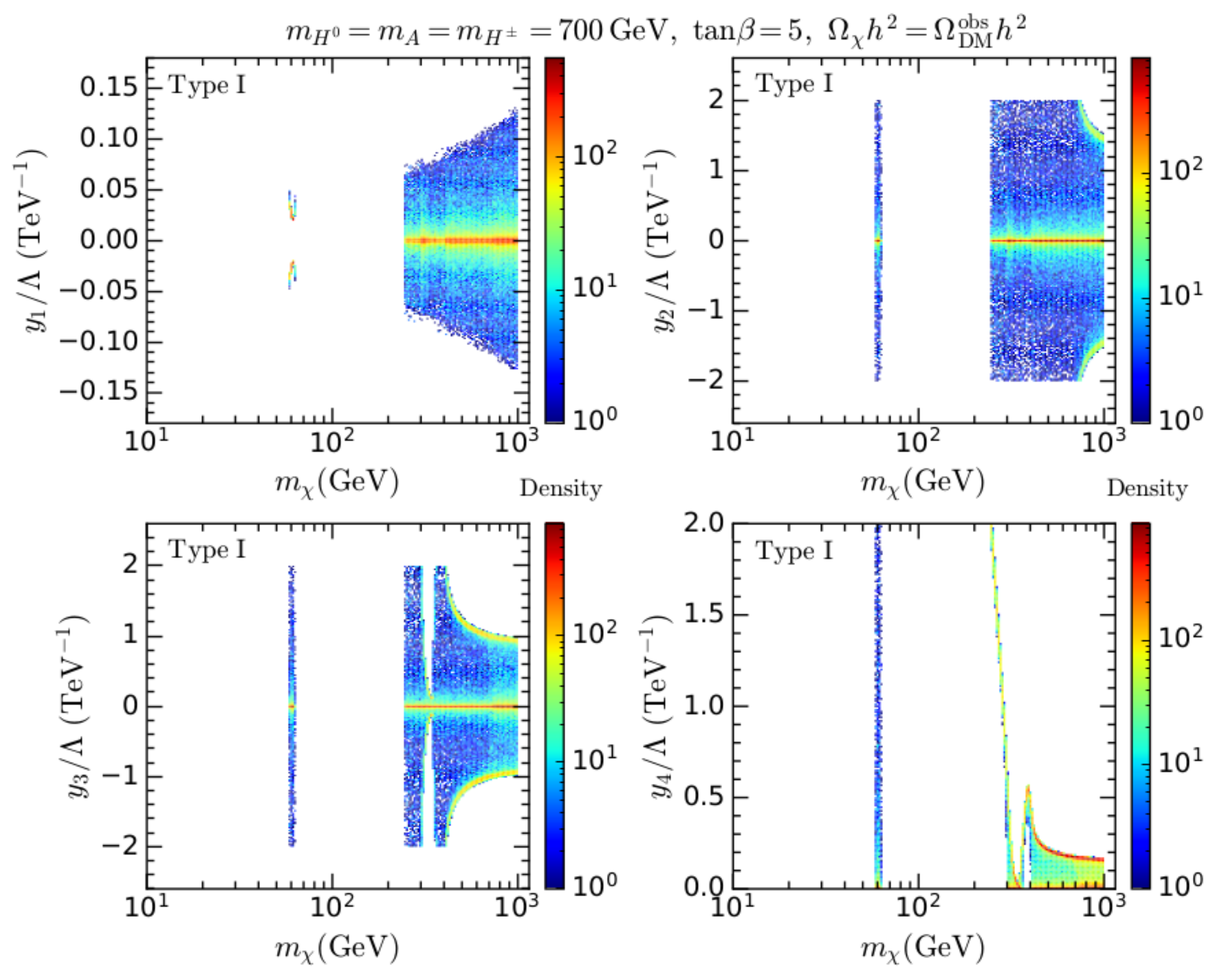}
    \caption{The same as Fig.~\ref{fig:type1_bchm1} for the Type I 2HDM with $\tan\beta=5$ and $\mH=\mA=\mHc=700\GeV$ (benchmark 3 in Table~\ref{tab:benchmarks}). }
    \label{fig:type1_bchm3}
\end{figure}

\clearpage

\begin{figure}[h]
    \centering
    \includegraphics[width=0.725\textwidth]{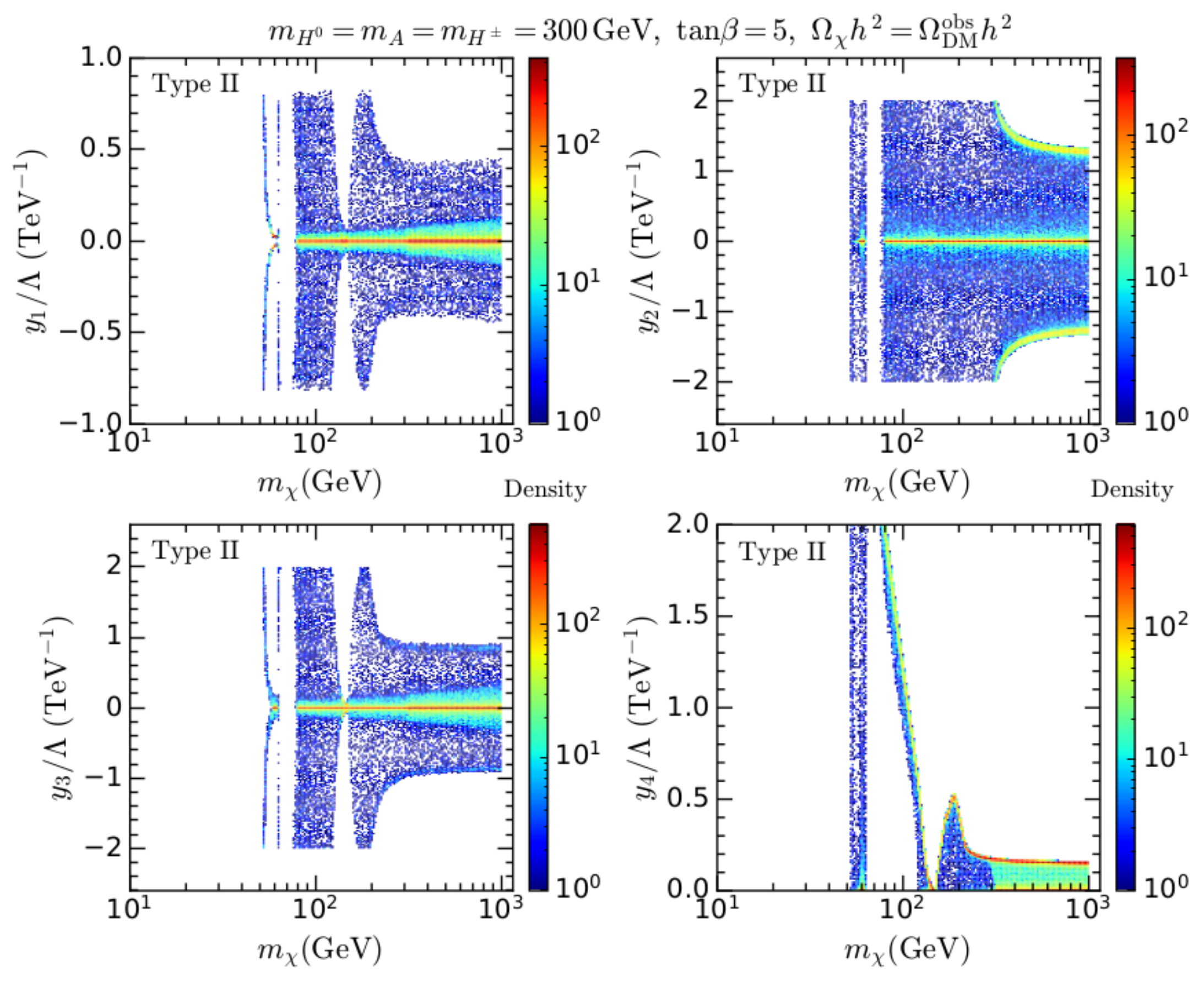}
    \caption{The same as Fig.~\ref{fig:type1_bchm1} for the Type II 2HDM. }
    \label{fig:type2_bchm1}
\end{figure}

\begin{figure}[h]
    \centering
    \includegraphics[width=0.725\textwidth]{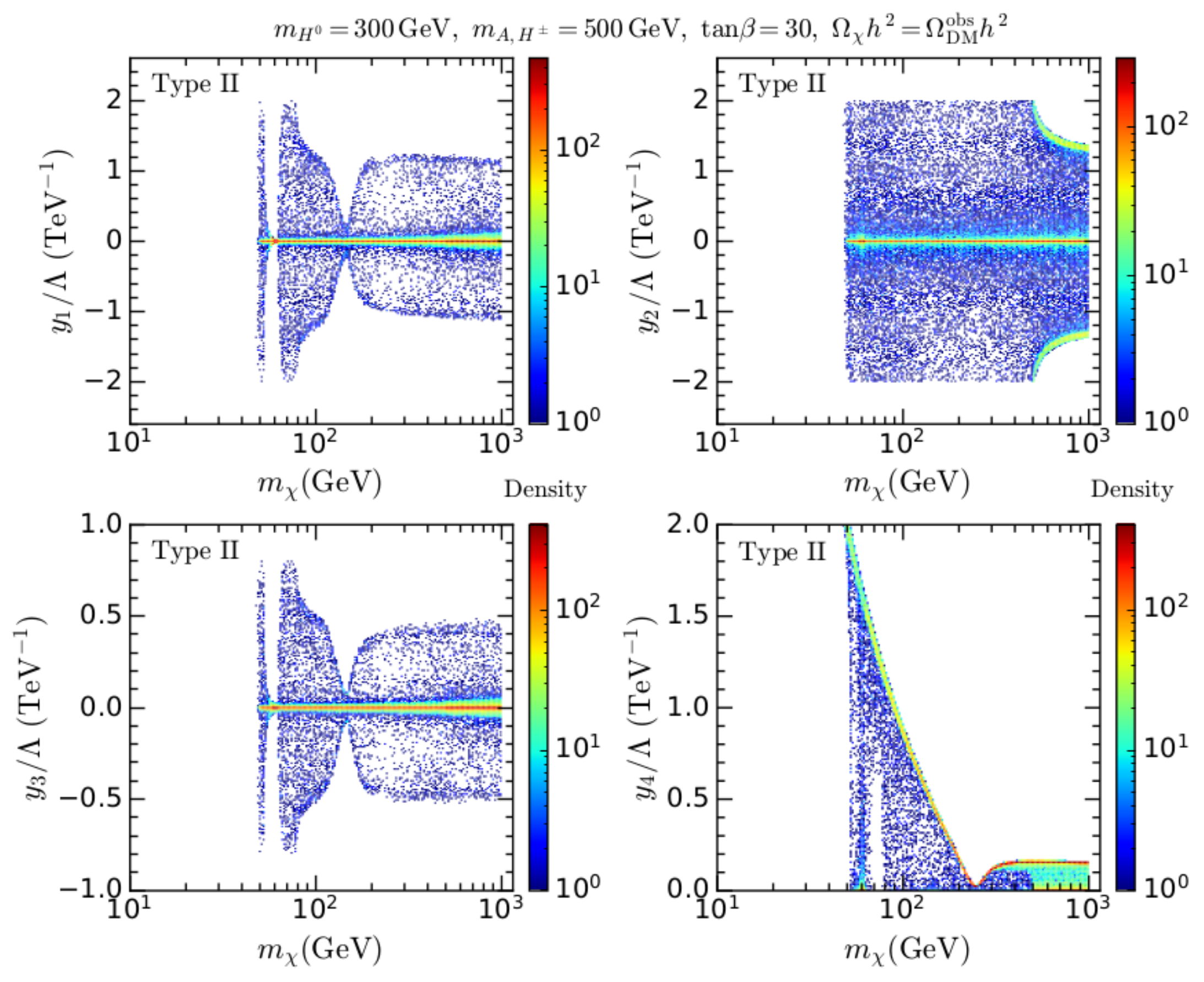}
    \caption{The same as Fig.~\ref{fig:type1_bchm1} for the Type II 2HDM with $\tan\beta=30$, $\mH=300\GeV$ and $\mA=\mHc=500\GeV$ (benchmark 2 in Table~\ref{tab:benchmarks}).}
    \label{fig:type2_bchm2}
\end{figure}

The other two benchmarks for the Type I 2HDM present similar structures, as shown in Figs.~\ref{fig:type1_bchm2} and \ref{fig:type1_bchm3}. The main differences are due to the positions of the various funnels and thresholds discussed above. It is clear that, even for quite large masses of the extra Higgs states, large regions of the DM mass below 1 TeV are rescued.

\begin{figure}[h]
    \centering
    \includegraphics[width=0.725\textwidth]{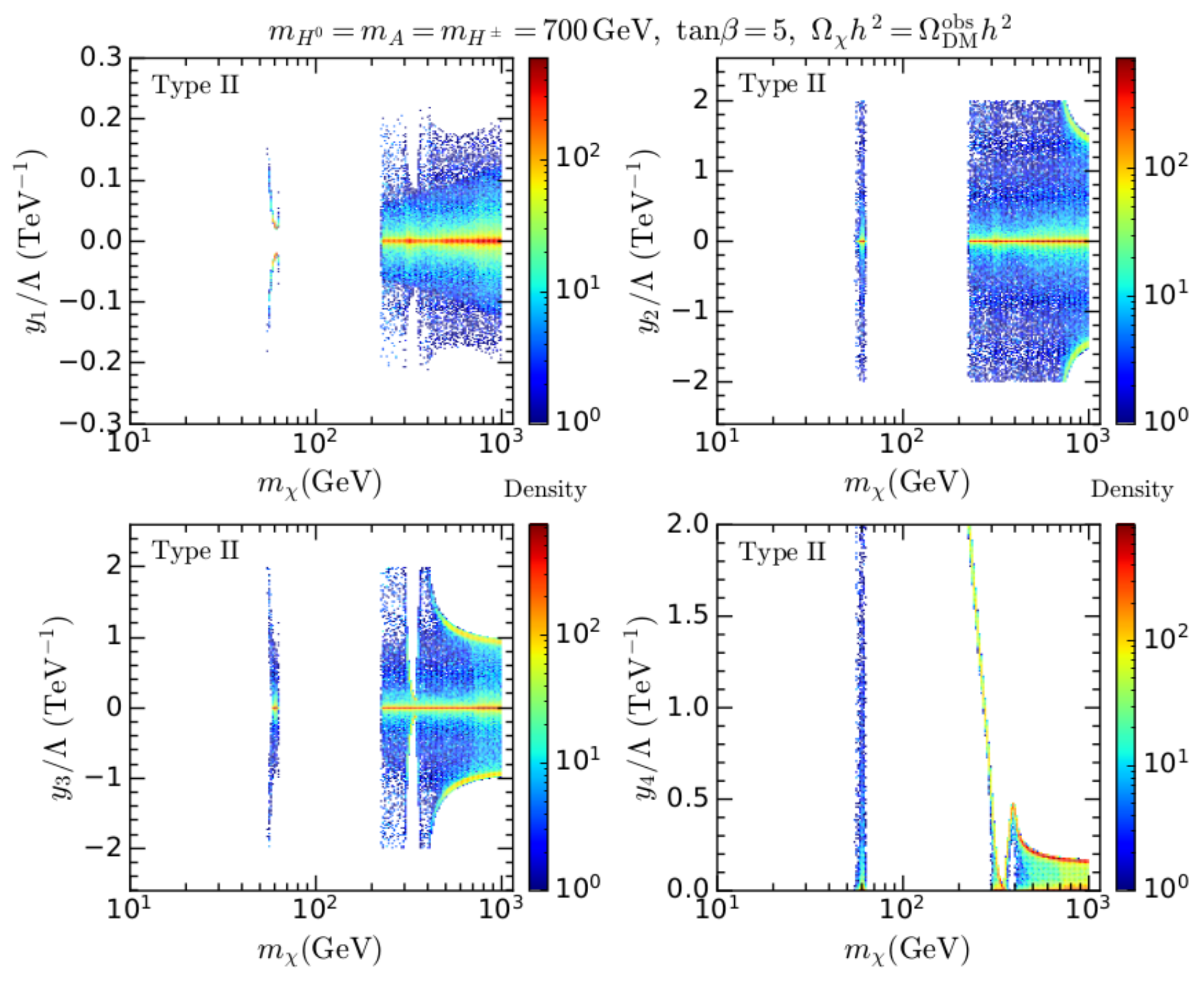}
    \caption{The same as Fig.~\ref{fig:type1_bchm3} for the Type II 2HDM. }    
    \label{fig:type2_bchm3}
\end{figure}

Even more interesting is the behavior of the Type II 2HDM. The funnels are still in the same places but the continuum of allowed values of $m_\chi$ typically starts
much sooner, quite below 100 GeV.
The reason is that for the Type II the coupling of the pseudoscalar to $d-$quarks is enhanced by $\tan \beta$, while for the Type I is suppressed by $\cot\beta$, see Table~\ref{tab:couplings}. Hence, for $\tan\beta=5$, the annihilation cross section mediated by $A$ in the  $s-$channel is ${\cal O}(500)$ times larger than for the Type I. 
The coupling of the heavy Higgs  to \emph{d}-quarks is also enhanced by $\tan\beta$, which makes the blind spot cancellation (Eq.~\ref{BN}) feasible even if $m_{H^0}$ is much larger than 125 GeV. Consequently, the acceptable values of $y_1$ can be much larger in the Type II than in the Type I. 
All this is shown in Figs.~\ref{fig:type2_bchm1}, \ref{fig:type2_bchm2} and \ref{fig:type2_bchm3} 
for the three Type II benchmarks listed in Table~\ref{tab:benchmarks}.

\begin{figure}[t]
    \centering
    \includegraphics[width=0.49\textwidth]{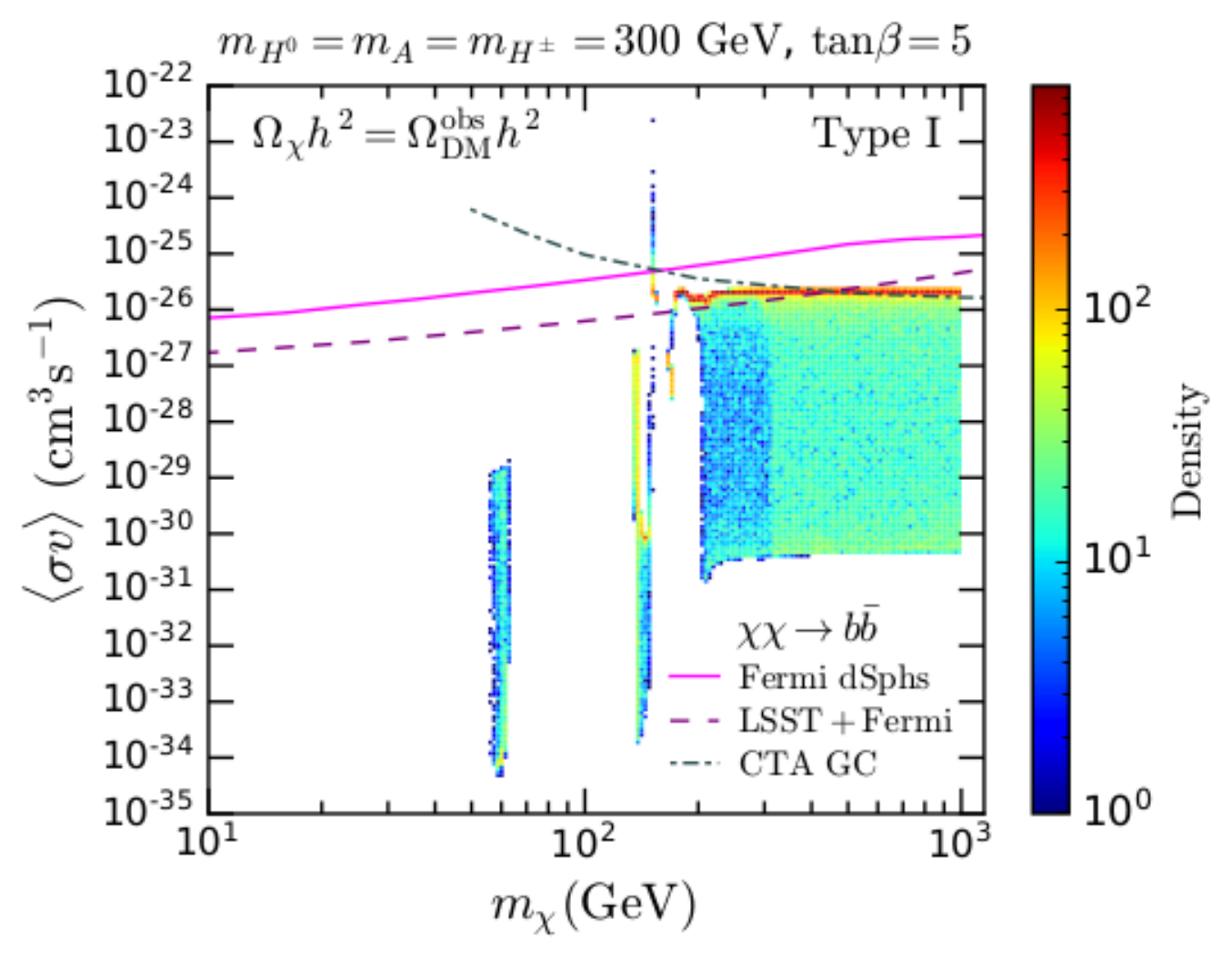}
    \includegraphics[width=0.49\textwidth]{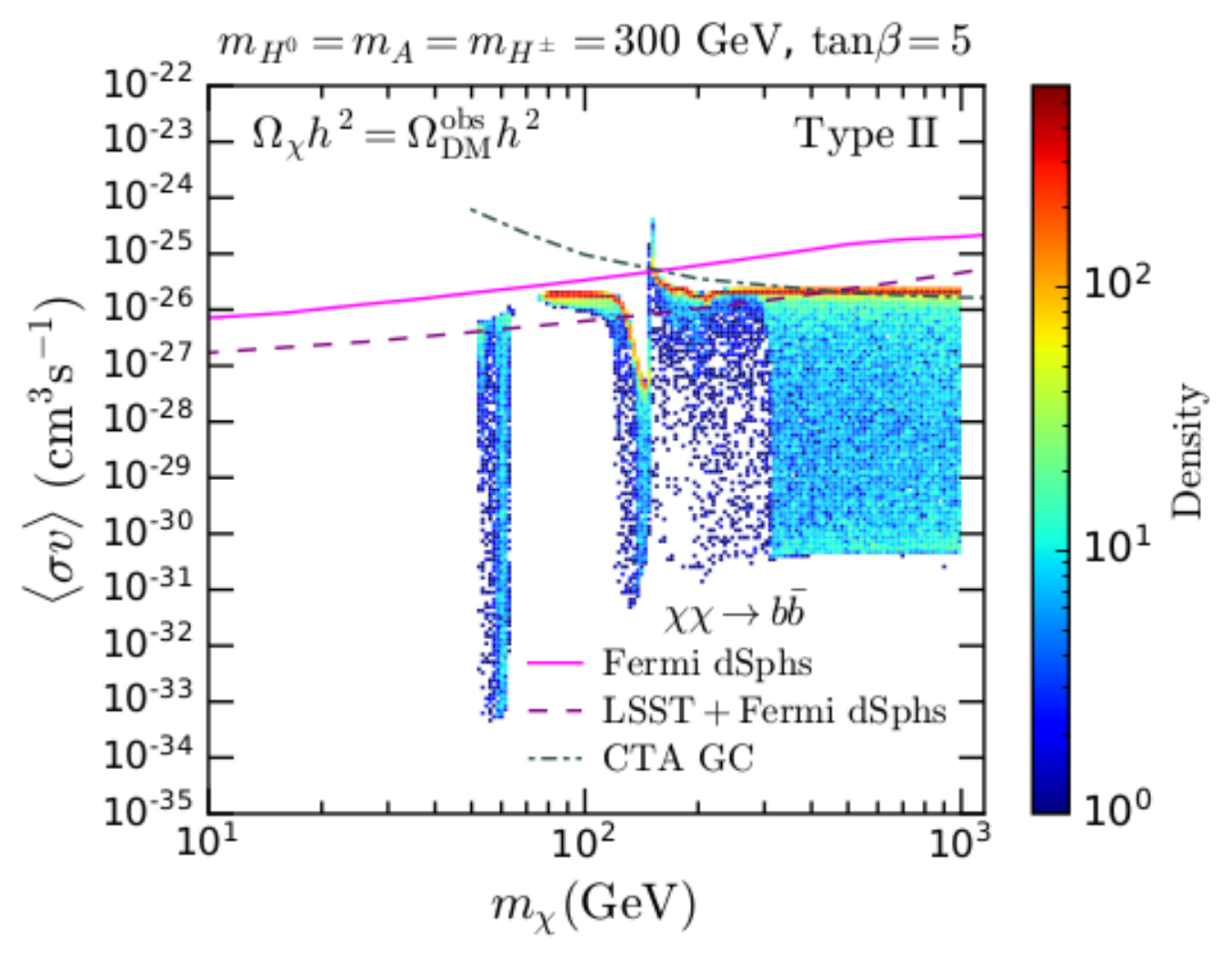}    
    \caption{Thermally averaged annihilation cross section as a function of the DM mass for the Type I (left) and Type II (right), benchmark model I. We also show the upper bound on $\langle \sigma v\rangle$ for the $b\bar{b}$ annihilation channel from Fermi LAT observations (solid magenta) of dwarf spheroidal galaxies (dSphs) \cite{Fermi-LAT:2016uux} and sensitivity projections from LSST + Fermi LAT dSph data (dashed purple) \cite{Drlica-Wagner:2019xan} and the  expected sensitivity of CTA to DM annihilation in the Galactic center (GC) for the same annihilation channel (dot-dashed dark grey) \cite{Acharyya:2020sbj}. }
    \label{fig:sigmav}
\end{figure}

Let us now discuss prospects of indirect detection. 
In Fig.~\ref{fig:sigmav}, we show the predictions for the thermally averaged annihilation cross section at zero velocity for the points in Figs.~\ref{fig:type1_bchm1} and \ref{fig:type2_bchm1} for the Type I (left) and II (right) 2HDM, respectively, which correspond to  the benchmark 1 in Table~\ref{tab:benchmarks}. We observe that in the region around the SM-like Higgs funnel, where DM annihilates primarily in the $b\bar{b}$ channel, the points allowed by DD are below the bounds from null searches with gamma rays in dwarf spheroidal galaxies (dSphs) by the Fermi-LAT Collaboration~\cite{Fermi-LAT:2016uux} (solid magenta). For the Type I, these points are even well below the sensitivity projection obtained combining  discoveries of new dSphs with the upcoming Large Synoptic Survey Telescope (LSST) with continued Fermi-LAT observations~\cite{Drlica-Wagner:2019xan} (dashed purple). For the Type II, on the other hand, LSST + Fermi-LAT data could probe  a small corner of the parameter space around the Higgs  funnel. The $b\bar{b}$ channel also dominates the resonant annihilation via the pseudoscalar and the  heavy Higgs around  $m_\chi\sim m_A/2, m_{H^0}/2$ ($=150\GeV$ for the benchmark 1). In this region the Fermi-LAT instrument excludes very few points of the parameter space  and LSST  + Fermi-LAT could probe a slightly wider region than current dSphs gamma-ray observations. The Cherenkov Telescope Array (CTA) (dot-dashed dark grey), on the other hand, would not be able to shed light on this kind of models, since its projected sensitivity to the region where DM annihilates mainly into SM channels is comparable to the leading bounds from the Milky Way satellites, recall that annihilation into at least one Higgs in the final state dominates the $m_\chi\gtrsim200\GeV$ region. Similar conclusions can be extracted from the  remaining benchmark models.

\clearpage
\section{Conclusions}
\label{sec:Conclusions}

Higgs-portal scenarios are among the most economical frameworks for dark matter (DM). Unfortunately, its simplest version, consisting of the Standard Model (SM) content plus a neutral particle, is essentially excluded by direct detection (DD) bounds. The main reason is that, in such scenario, the DM annihilation  in the early Universe (and thus the relic abundance, $\Omega_{\rm DM} h^2$) and the DD cross section are controlled by
the same coupling.

In this paper we have considered a slightly modified scenario, where the DM sector still consists of a single particle (a neutral fermion), but the Higgs sector comprises two Higgs doublets, i.e. the well-known 2HDM. The main virtue of this framework is that the previous one-to-one relationship between the DM relic abundance and the DD cross section does not hold anymore, thus giving rise to vast allowed regions in the parameter space.

The minimal scenario entails to consider an effective-theory Lagrangian (see Eqs.~\ref{L1} and~\ref{L3}) with four independent couplings, plus the masses of the three extra Higgs-states. Experimental constraints require that the properties of the light Higgs boson are quite close to those of the SM. This is known as the alignment limit, which we have assumed throughout the analysis. Such limit can be accomplished either because the extra Higgs states are very heavy (alignment from decoupling) or because the parameters in the scalar potential are aligned in an appropriate way. We have analyzed both cases. Actually, the former is indistinguishable from the ordinary Higgs-portal, hence our results simply illustrate (and up-date) the fact that almost its entire parameter-space is excluded by DD.

The second case (alignment without decoupling) is far more interesting. The four independent DM couplings are relevant for DM annihilation, but not all of them are equally important for DD processes (actually, one of them is almost totally irrelevant). This fact opens enormously the allowed parameter space.
In addition, the DM couplings that are relevant for DD can have opposite-sign contributions to the total cross-section, leading to a possible cancellation that does not occur for DM annihilation. These are the so-called blind spots (or blind regions). Finally, as it is well-known, the presence of the extra Higgs states enables new paths for resonant DM annihilation ($A-$funnel and $H^0-$funnel), which are safe regarding DD. 

We have illustrated the previous facts by analyzing in detail several benchmark points for the Type I and Type II 2HDMs (the results for the X and Y 2HDMs are equivalent). The main difference between them is due to  the different couplings of the extra Higgs states to the SM fermions, e.g. the coupling of the pseudoscalar, $A$, to $d-$quarks is enhanced (suppressed) by $\tan\beta$ ($\cot\beta$) for the Type II (Type I) 2HDM. This makes the DM annihilation mediated by $A$ in the  $s$-channel much more efficient for the Type II, thereby enhancing its allowed parameter space, specially for light DM.

Finally, we have discussed the prospects for indirect detection. At present, essentially all the parameter space consistent with the relic abundance and DD bounds, is also safe with respect to indirect detection. However, some regions in the parameter space of the Type II 2HDM could be probed by LSST + Fermi LAT dwarf spheroidal galaxies data.

\section*{Acknowledgments}
The work of JAC has been partially supported by Spanish Agencia Estatal de Investigaci\'on through the grant ``IFT Centro de Excelencia Severo Ochoa SEV-2016-0597’', and by MINECO project FPA 2016-78022-P and MICINN project PID2019-110058GB-C22.
The work of AD was partially supported by the National Science Foundation under grant PHY-1820860.
SR was supported by the Australian Research Council.

\bibliographystyle{JHEP}  
\bibliography{references}

\end{document}